\documentclass[twocolumn,showpacs,amsmath,amssymb]{revtex4}

\newcommand{\beq}{\begin{equation}}
\newcommand{\eeq}{\end{equation}}
\newcommand{\bea}{\begin{eqnarray}}
\newcommand{\eea}{\end{eqnarray}}

\usepackage{latexsym}
\usepackage{amssymb}
\usepackage{amsfonts}
\usepackage{amsmath}
\usepackage[dvips]{graphicx}

\begin{document}

\title{Numerical analysis of backreaction in acoustic black holes}

\author{Roberto Balbinot$^{1}$, Alessandro Fabbri$^{2}$, Serena Fagnocchi$^{3,1}$ and Alessandro Nagar$^{4}$ }

\affiliation{$^{1}$ Dipartimento di Fisica, Universit\`a di Bologna and INFN sez. di Bologna,
                    Via Irnerio 46, 40126 Bologna Italy \\
   $^2$ Departamento de Fisica Teorica and IFIC, Centro Mixto Universidad de Valencia--CSIC. \\Facultad de Fis\`ica, Universidad de Valencia, 46100 Burjassot (Valencia), Spain\\
	 $^3$ Centro Enrico Fermi, Compendio Viminale, 00184 Roma, Italy \\
  $^{4}$ Dipartimento di Fisica, Politecnico di Torino and INFN sez. di Torino,
                    Corso Duca degli Abruzzi 24, 10129 Torino, Italy.}

\date{\today}

\begin{abstract}
	Using methods  of  Quantum  Field Theory in  curved spacetime, the  first order in $\hbar$ 
	quantum corrections to the  motion of a  fluid in an acoustic black hole configuration 
	are numerically computed. These  corrections arise  from the  non linear backreaction 
	of the emitted phonons. Time dependent (isolated system) and equilibrium configurations 
	(hole in a sonic cavity) are both analyzed.
\end{abstract}

\pacs{04.62.+v, 04.70.Dy, 47.40.Ki}

\maketitle

\section{Introduction}
Black hole radiation predicted by Hawking in 1974 \cite{hawking} is one of the most 
spectacular results of modern theoretical physics.

Even more surprising is the fact that this effect is not peculiar of gravitational 
physics, but is also expected in many completely different contexts of condensed matter 
physics \cite{unruh81,libro,BLV}. A fluid undergoing supersonic motion is the simplest 
example of what one calls an ``acoustic black hole''. For this configuration 
Unruh~\cite{unruh81}, using Hawking arguments, predicted an emission of thermal phonons. 
This emission affects the behaviour of the underlying fluid because of the non linearity 
of the hydrodynamical equations governing its motion.

Using methods borrowed from Quantum Field Theory in curved spacetime, this quantum 
backreaction has been studied for the first time in~\cite{nostri}, where the 
 the first order in $\hbar$ corrections to the classical hydrodynamical equations were given.
Because of intrinsic mathematical difficulties, the analysis was 
restricted to the region very close to the ``sonic horizon'' of the acoustic black hole; 
i.e., the region where the fluid motion changes from subsonic to supersonic. There, analytical 
expressions for the quantum corrections to the density and velocity of the mean flow have 
been provided.

However, to  have a detailed description throughout the entire system one has to proceed 
with numerics. This will be the aim of our present paper, which is organized as follows: in 
Sec.~\ref{sec:acousticBH} we outline the classical fluid configuration which describes an 
acoustic black hole; the quantum backreaction equations are discussed in Sec.~\ref{sec:qbckr}, 
with emphasis on the choice of quantum state in which the phonons field has to be quantized. 
In Secs.~\ref{sec:U} and~\ref{sec:HH} we give the numerical estimates for the quantum correction 
to the mean flow in two different cases: isolated system and system in equilibrium 
in a sonic cavity respectively. 
Section~\ref{sec:conclusions} contains the final discussion. 

\section{The acoustic black hole}
\label{sec:acousticBH}

An acoustic black hole is a region of a fluid 
where its motion is supersonic. Here sound can not escape upstream being dragged by 
the fluid. The boundary of this region is formed by sonic points where the speed of 
the fluid equals the local speed of sound. This is the acoustic horizon. A simple 
device to establish a transonic flow is a converging diverging 
de Laval nozzle~\cite{courant49,libro}. For stationary free fluid flow the acoustic 
horizon occurs exactly at the waist of the nozzle.

The basic equations describing the system at the classical level are the continuity and the Bernoulli 
equations. We assume a one-dimensional stationary flow, therefore all the relevant quantities depend 
on $z$ only, the spatial coordinate running along the axis of the de Laval nozzle. The continuity 
equation then reads
\beq
\label{continuity}
A(z)\rho(z) v(z)=\mbox{const}=D \ ,
\eeq
where $A$ is the area of the transverse section of the nozzle, $\rho$ 
the fluid density and $v$ the fluid velocity. The Bernoulli equation, 
under the above hypothesis, gives
\beq
\label{bernoulli}
\frac{v^2}{2}+\mu(\rho)=0 \ ,
\eeq
where $\mu(\rho)$ is the enthalpy. We have further assumed the fluid to be homentropic and irrotational. 
The speed of sound $c$ is defined as
\beq
\label{c.def}
c^2=\rho\frac{d\mu}{d\rho} \ .
\eeq
For constant $c$ (the case we consider) integration of (\ref{c.def}) gives
\beq
\mu(\rho)=c^2 \ln \frac{\rho}{\rho_0} \ ,
\eeq
which inserted in Bernoulli equation yields
\beq
\label{rho.profile}
\rho=\rho_0 e^{-\frac{v^2}{2c^2}} \ ,
\eeq
with $\rho_0$ a constant. The assumed constancy of the speed of sound also gives the pressure 
$p$ as $p=c^2 \rho$. The velocity profile describing the acoustic black hole is chosen as
\beq 
\label{v.profile}
v=c\left\{ \frac{2}{\pi}\arctan [\beta(z-z_H)]-1\right\}\ ,
\eeq
\begin{figure}[t]
\begin{center}
\includegraphics[width=85mm , height=80 mm]{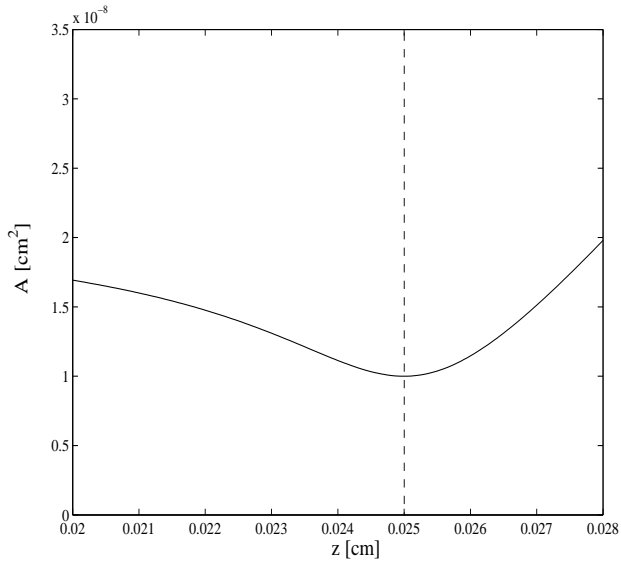}
\caption{Dependence of the cross section of the de Laval nozzle on the position $z$ for
a velocity field given by Eq.~(\ref{v.profile}) and depicted in Fig.~\ref{label:fig3}.
The vertical dashed line corresponds to the location of the sonic horizon $z_H=0.025$ cm.}
\label{label:fig2}
\end{center}
\end{figure}
where $z=z_H$ denotes the position of the waist of the nozzle (the sonic horizon). 
In the laboratory frame the fluid is moving from right to left, so $v<0$ and the 
sonic horizon occurs where $v=-c$. The constant $D$ entering the continuity equation 
is determined by requiring the fluid to be sonic at the waist; i.e.,
\beq
D=-c A_H \rho_0 e^{-1/2}=-\frac{A_H p_H}{c} \ ,
\eeq
where $A_H$ is the area at the horizon, $p_H$ the pressure and $\rho_0=p_H e^{1/2}/c^2$. 
Given this, the profile of the nozzle can be computed from Eq.~(\ref{continuity}) and is depicted 
in Fig.~\ref{label:fig2}, where we have used $A_H=10^{-8}$ cm$^{2}$, $\beta=600$ cm$^{-1}$, 
$p_H=2\times 10^6$ Pa and $c=250$ m/s. 
These latter two are typical values for liquid Helium. The profiles of density and velocity are 
shown in Fig.~\ref{label:fig3}, where the significant range of $z$ is $[0, 0.05]$ cm; the horizon 
lies at $z_H=0.025$ cm and its location is indicated by a vertical dashed line in the figures. 
In the region $z>z_H$ the motion of the fluid is subsonic; the acoustic black hole is the region $z<z_H$.

As shown by Unruh~\cite{unruh81}, sound  waves  propagating in an inhomogeneous fluid are 
described as a massless scalar field propagating  in an effective  curved  spacetime  described 
by an ``acoustic metric''  which depends on $\rho$ and $v$. Quantization of these  modes leads 
to the conclusion  that in  presence of a sonic horizon a thermal emission of phonons is expected, 
in complete analogy of what Hawking found for gravitational black holes. The emission temperature 
of the phonons is $ T_H=\hbar k/(2\pi c \kappa_B)$, where $k$ is the surface gravity of the sonic 
horizon, defined as
\beq
\label{eq:k}
k=\left. \frac{1}{2}\frac{d}{dn}(c^2-v^2)\right|_{z_H} \ .
\eeq
$\kappa_B$ is the Boltzmann constant and $n$ is the normal to the horizon.
\begin{figure}[t]
\begin{center}
\includegraphics[width=80mm , height=75 mm]{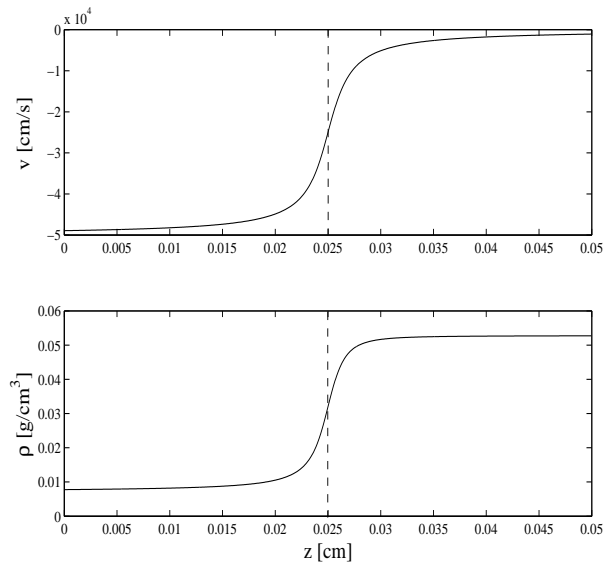} \\
\caption{Velocity ({\it top}) and density ({\it bottom}) from Eq.~(\ref{v.profile}). The vertical dashed
line correspond to the location of the sonic horizon ($z_{H}=0.025$ cm). The sonic black hole corresponds 
to $z<z_{H}$.}
\label{label:fig3}
\end{center}
\end{figure}
For the specific acoustic black hole model we consider in this 
paper, $T_H=1.1598\times 10^{-5}$ $^\circ$K.

\begin{figure*}[t]
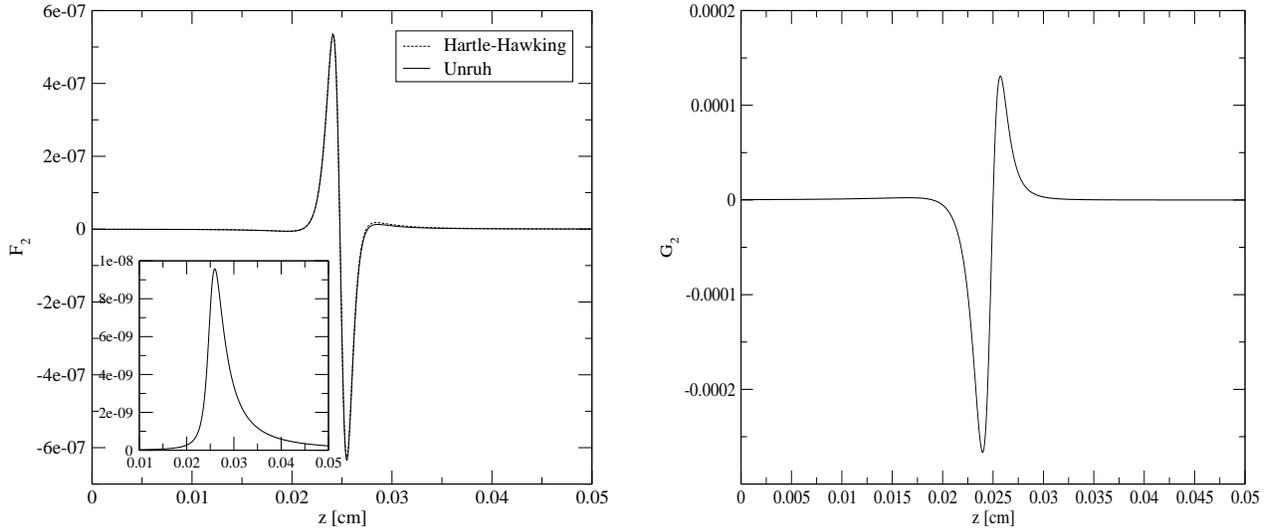

\begin{center}
\includegraphics[width=80mm , height=70 mm]{fig3a.eps}\qquad 
\includegraphics[width=80mm , height=70 
mm]{fig3b.eps}
\caption{Quantum sources in the Unruh state (solid lines) and in the Hartle-Hawking state (dashed line). 
To appreciate the difference in $F_2$ taken in the Hartle-Hawking and Unruh states, we show it in the
inset. The $G_2$ is the same for both states. The sonic horizon is located at $z_H=0.025$ cm.}
\label{label:fig4}
\end{center}
\end{figure*}
\section{The backrection equations}
\label{sec:qbckr}

The phonons quantum emission previously discussed modifies the underlying fluid flow according to 
the backreaction equations derived in Ref.~\cite{nostri}, to which we refer for further details.
For a one-dimensional flow, they read
\begin{align}
\label{back1}
A\rho_B+\partial_z(A\rho_B v_B)&=\partial_z\left[\frac{1}{c}(\langle T_{tz}^{(2)}\rangle +v\langle T_{zz}^{(2)}\rangle )\right] \ , \\
\label{back2}
A\left[ \dot{\psi}_B+\frac{v_B^2}{2}+\mu(\rho_B) \right]&=\frac{1}{2}\langle T^{(2)}\rangle\ .
\end{align}
Here $\rho_B$ and $v_B$ are the quantum corrected density and velocity fields and $\psi_B$ is the velocity potential; 
i.e., $\partial_z \psi_B=v_B$; the overdot stands for time derivative. The $\langle T_{ab}^{(2)}\rangle$ which 
drive the backreaction are the quantum expectation values of the pseudo energy momentum tensor quadratic in the 
phonons field. To evaluate $(\rho_B,v_B)$ up to $O(\hbar)$ terms, the r.h.s. of the backreaction equations 
(\ref{back1}) and (\ref{back2}) needs just to be evaluated on the classical background $(\rho,v)$ of Sec.~\ref{sec:acousticBH}.

The quantum state of the field in which the expectation values have to be computed depends on the physical situation 
one wants to describe. For an isolated hole, the escaping phonons radiation leads to a time variation of the 
underlying medium, i.e. $\rho_B(t,z)$ and $v_B(t,z)$. The appropriate quantum state in this case is the analogue 
of the Unruh state~\cite{unruh76}.  In case the system is maintained in thermal equilibrium with the surroundings 
(that is, putting a sonic cavity in the subsonic asymptotic  right region), the quantum state is the analogue of the Hartle-Hawking state~\cite{HH}, 
the thermal equilibrium state at $T=T_H$. In this case the system remains stationary, i.e. $\rho_B(z)$ and $v_B(z)$. 

\begin{figure*}[t]
\begin{center}
\includegraphics[width=80mm ,height=70mm]{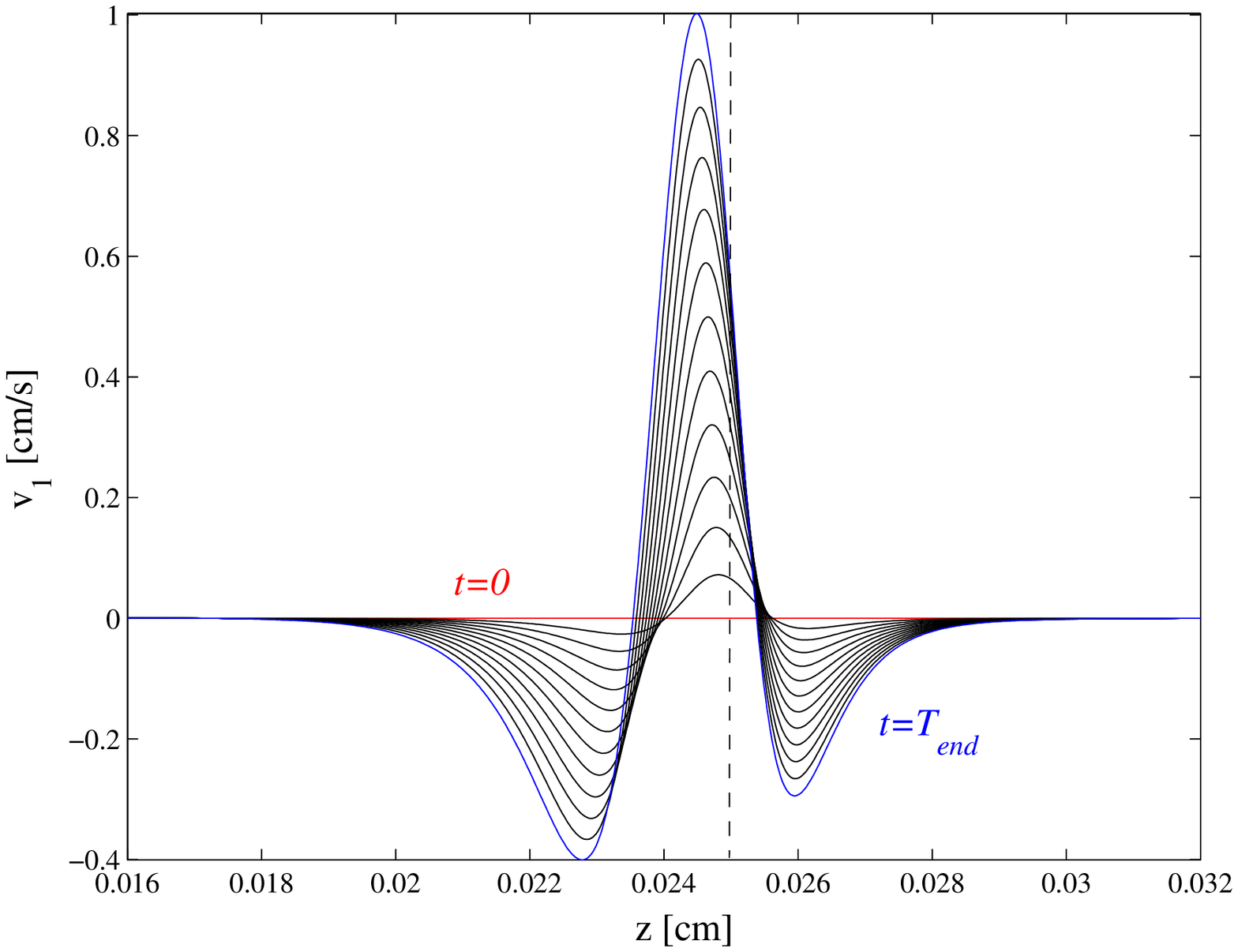}\qquad
\includegraphics[width=80mm , height=72 mm]{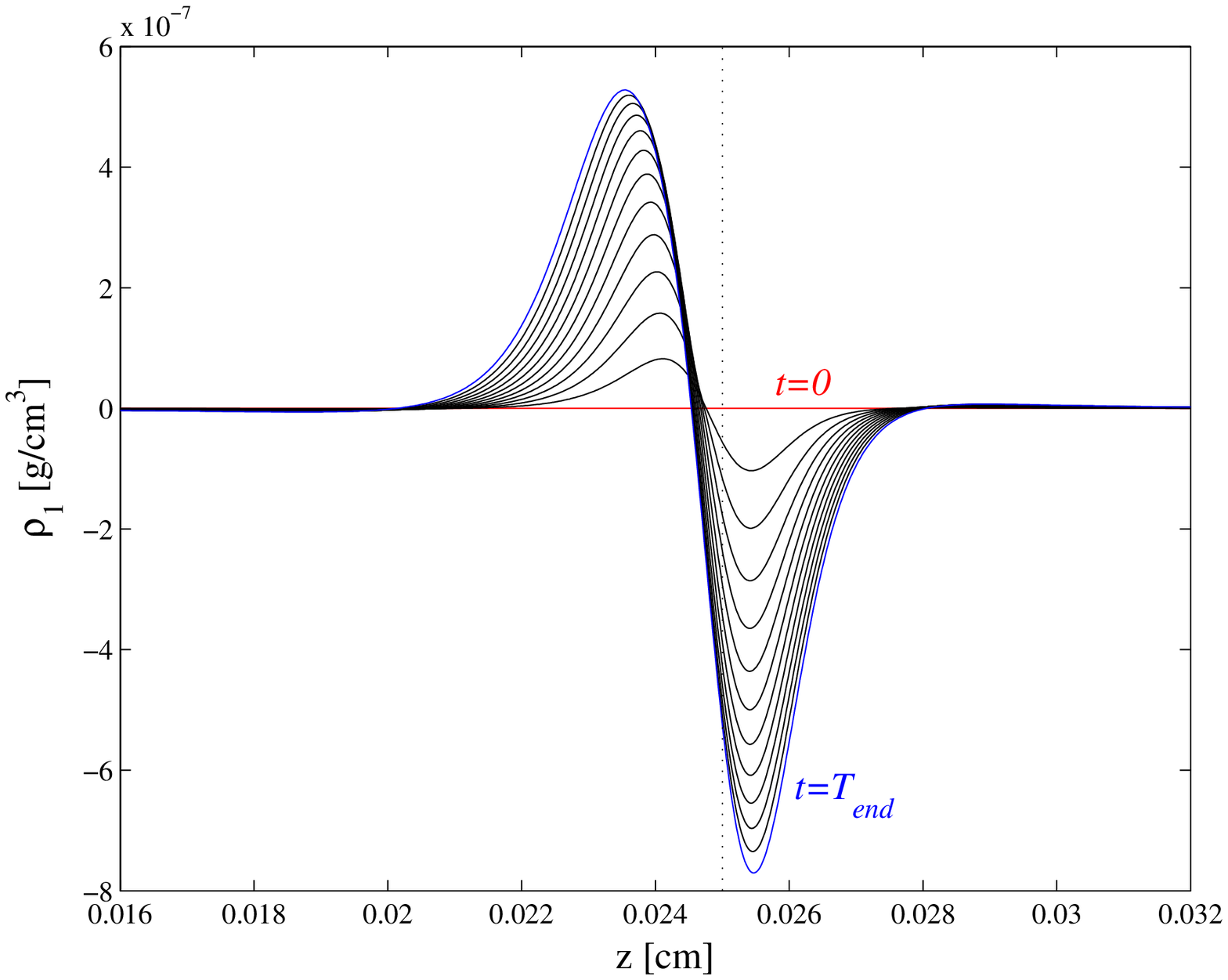}
\caption{Unruh state: time evolution of the backreaction equations for $c\kappa t\ll 1$. Snapshots of the 
quantum correction to the velocity $v_1$ ({\it left panel}) and to the density ({\it right panel}) for an
evolution time $t_{\rm end}=0.2\,t_{\rm max}\approx2.09\times 10^{-2}$ $\mu$s ($c\kappa t=0.2$). The vertical
dashed line corresponds to the location of the classical sonic horizon. The delay between one snapshot and
the other (between $t=0$ and $t=t_{\rm end}$) is $\Delta t\approx 1.74\times 10^{-3}$ $\mu$s.}
\label{label:fig5}
\end{center}
\end{figure*}

Neglecting backscattering of the phonons, $\langle T_{ab}^{(2)}\rangle$ can be approximated with the Polyakov stress 
tensor~\cite{polyakov, sandro-pepe}. Introducing for the sake of simplicity null coordinates
\beq
x_\pm=c\left( t\mp \int \frac{dz}{c\pm v}\right)\,
\eeq
the Polyakov stress tensor reads: 
\bea
\langle T_{\pm\pm}^{(2)}\rangle &=& -\frac{\hbar}{12\pi}C^{1/2}C^{-1/2}_{,\pm\pm}+\Delta_{\pm\pm}\\
\langle T^{(2)}\rangle          &=& \frac{\hbar}{6\pi}C^{-1}\left(\ln C\right)_{,+-} \ ,
\eea
where
\beq
C=\frac{\rho}{c}
\frac{c^2-v^2}{c^2}
\eeq
and $\Delta_{\pm\pm}$ are functions which depend on the choice of the quantum state of the phonons field.  
For the Unruh state:
\bea\label{delta++}
\Delta_{++}\equiv\Delta^{\rm U}_{++}&=&0 \ ,\\
\Delta_{--}\equiv\Delta^{\rm U}_{--}&=&\frac{\hbar k^2}{48\pi c^4}\ .
\eea
For the Hartle-Hawking  state instead
\beq\label{delta.pm}
\Delta_{\pm\pm}\equiv\Delta^{\rm HH}_{\pm\pm}=\frac{\hbar k^2}{48\pi c^4}\ .
\eeq

From Eqs.~(\ref{delta++})--(\ref{delta.pm}) it follows that, in the asymptotic subsonic region $z\rightarrow +\infty$, $\Delta^{\rm U}_{\pm\pm}$ 
describes a flux of phonons at a temperature $T_H$, whereas $\Delta^{\rm HH}_{\pm\pm}$ describes 
a two-dimensional gas of phonons at thermal equilibrium at the temperature $T_H$. To find first 
order in $\hbar$ corrections to the classical sonic black hole fluid configuration $(\rho(z),v(z))$ 
described in Eqs.~(\ref{rho.profile}) and~(\ref{v.profile}) we write
\bea
\psi_B&=&\psi(z)+\epsilon \psi_1(t,z)\ ,\\
\rho_B&=&\rho(z)+\epsilon \rho_1(t,z)\ ,
\eea
with $v_B=\partial_z \psi_B$ and $\epsilon$ is a dimensionless expansion parameter \cite{york}:
\beq
\epsilon=\frac{\hbar}{|D|A_H}\ . 
\eeq
For our system $\epsilon=1.317 \times 10^{-14}$. \\
The backreaction equations linearized in $\epsilon$ then become
\begin{align}
&\epsilon\bigg\{ A\dot{\rho}_1+\partial_z\left[A(\rho_1 v+\rho v_1)\right]\bigg\}\nonumber\\
&\qquad\qquad=c^2\partial_z\left[\dfrac{\langle T_{++}^{(2)}\rangle}{(c-v)^2}-\dfrac{\langle T_{--}^{(2)}\rangle}{(c+v)^2}\right]
\equiv \epsilon F_2 \ ,\label{eq.back.1}\\
&\epsilon\left[ A\left(\dot{\psi}_1+v v_1+\frac{c^2}{\rho}\rho_1\right)\right]=\frac{\langle T^{(2)}\rangle}{2}\equiv \epsilon G_2 \ .\label{eq.back.2}
\end{align}
Using the background equations (\ref{continuity}) and (\ref{bernoulli}), satisfied by 
$\rho$ and $v$, the continuity equation can be rewritten as
\beq
\label{continty.back.lin}
\dot{\rho_1}+v\rho_1'+\frac{v^2v'}{c^2}\rho_1-\dfrac{\rho v'}{v}\psi_1'+\rho\psi_1''=\frac{F_2}{A}\ ,
\eeq
whereas the Bernoulli equation is
\beq\label{bernoulli.back.lin}
\dot{\psi_1}+v\psi_1'+\frac{c^2}{\rho}\rho_1=\frac{G_2}{A}\ ,
\eeq
with a prime indicating derivative with respect to $z$. 

The profiles for the quantum sources $F_2$ and $G_2$ are depicted in 
Fig.~\ref{label:fig4} for the Unruh state (solid line) and for the Hartle-Hawking state 
(dashed line). The difference between the states is reflected on $F_2$ only (and is shown
in the inset in the left panel of the figure), while $G_2$, being related to the trace anomaly which is state-independent, is unchanged. One can note the 
appearance of a maximum and a minimum in the region $z\in [0.02,0.03]$ cm. Outside this 
range, $F_2$ and $G_2$ rapidly drop to zero. The analysis of Ref.~\cite{nostri}, being 
limited to the region very near to $z_H$, could not catch this non trivial structure.

\section{Unruh state}
\label{sec:U}

As said before, in Ref.~\cite{nostri} the backreaction equations were analytically 
solved just for $z\approx z_H$ to allow a Taylor expansion of the sources up to linear 
terms. In this section we compute the numerical solution all over the nozzle. 
We finite-difference the system of Eqs.~(\ref{continty.back.lin}) 
and~(\ref{bernoulli.back.lin}) and solve it numerically in the time domain as an initial 
value problem. The equations are discretized on an evenly spaced grid $(0,z_{\rm c})$ with
$z_{\rm c}=0.05$ cm. Following a standard convention in numerical fluid mechanics \cite{leveque}, 
we have used a staggered grid, i.e. both $z=0$ and $z=z_{\rm c}$ are thought to lie on cell 
{\it interfaces} while the hydrodynamics quantities are defined on cell {\it centers}. As a result, 
the first point of our computational domain is $z_1=\Delta z/2$ and the last is 
$z_{i_{\rm max}}=z_{\rm c}-\Delta z/2$. We notice that $\Delta z$ is chosen so that
the horizon is located at a cell interface. The reason for this is that, even though the square bracket on the r.h.s. of Eq. (\ref{eq.back.1})
is analytically regular at $z=z_{H}$ (where $v=-c$), the presence of the combination ($c+v$)
at denominator in Eq.~(\ref{continty.back.lin}) can give problems (i.e., division by zero) 
due to the discretization procedure. The use of a staggered grid bypasses this difficulty 
since the sonic point turns out to be always displaced  with respect to the grid points.
\begin{figure*}[t]
\begin{center}
\includegraphics[width=80mm , height=70 mm]{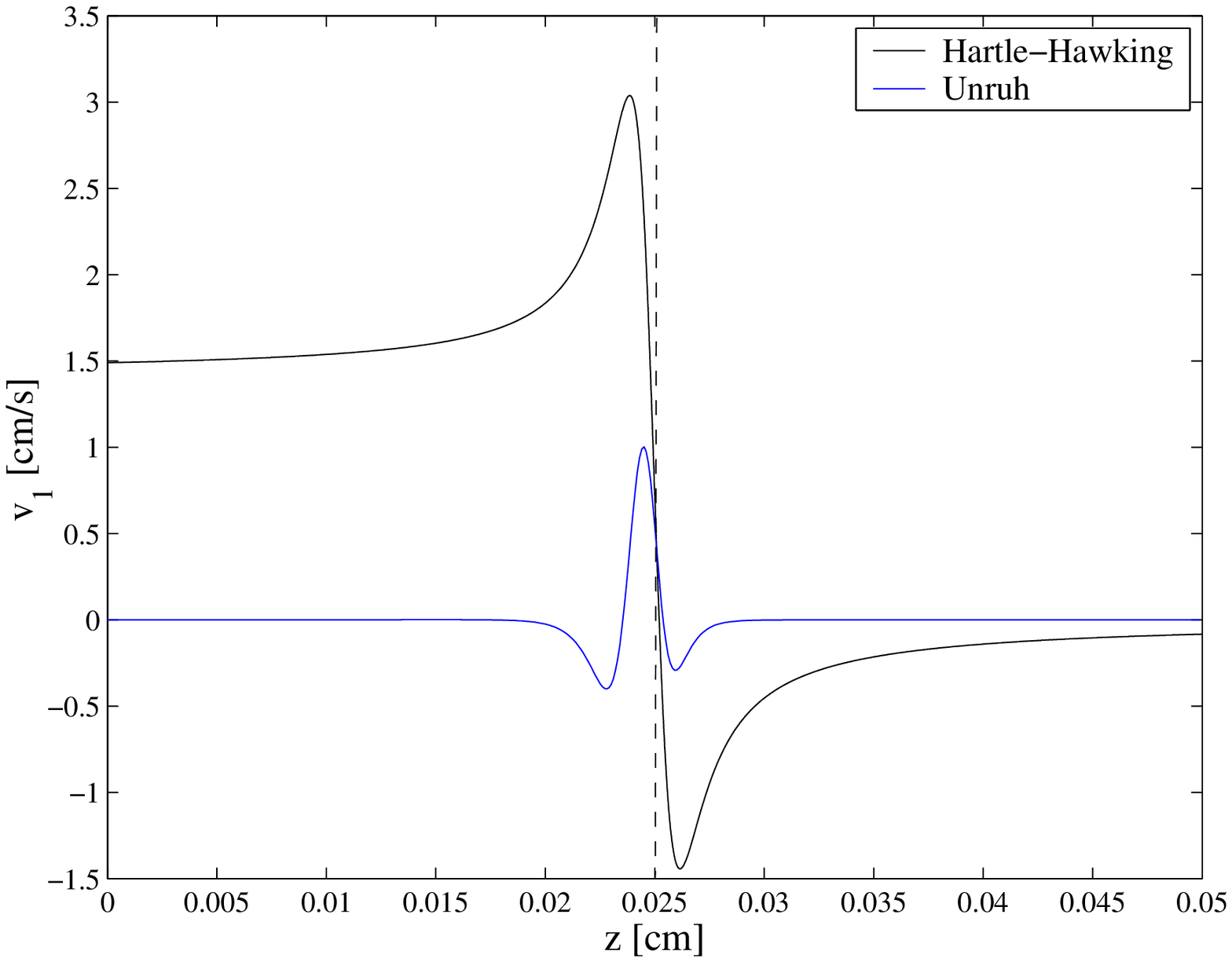}
\includegraphics[width=80mm , height=72 mm]{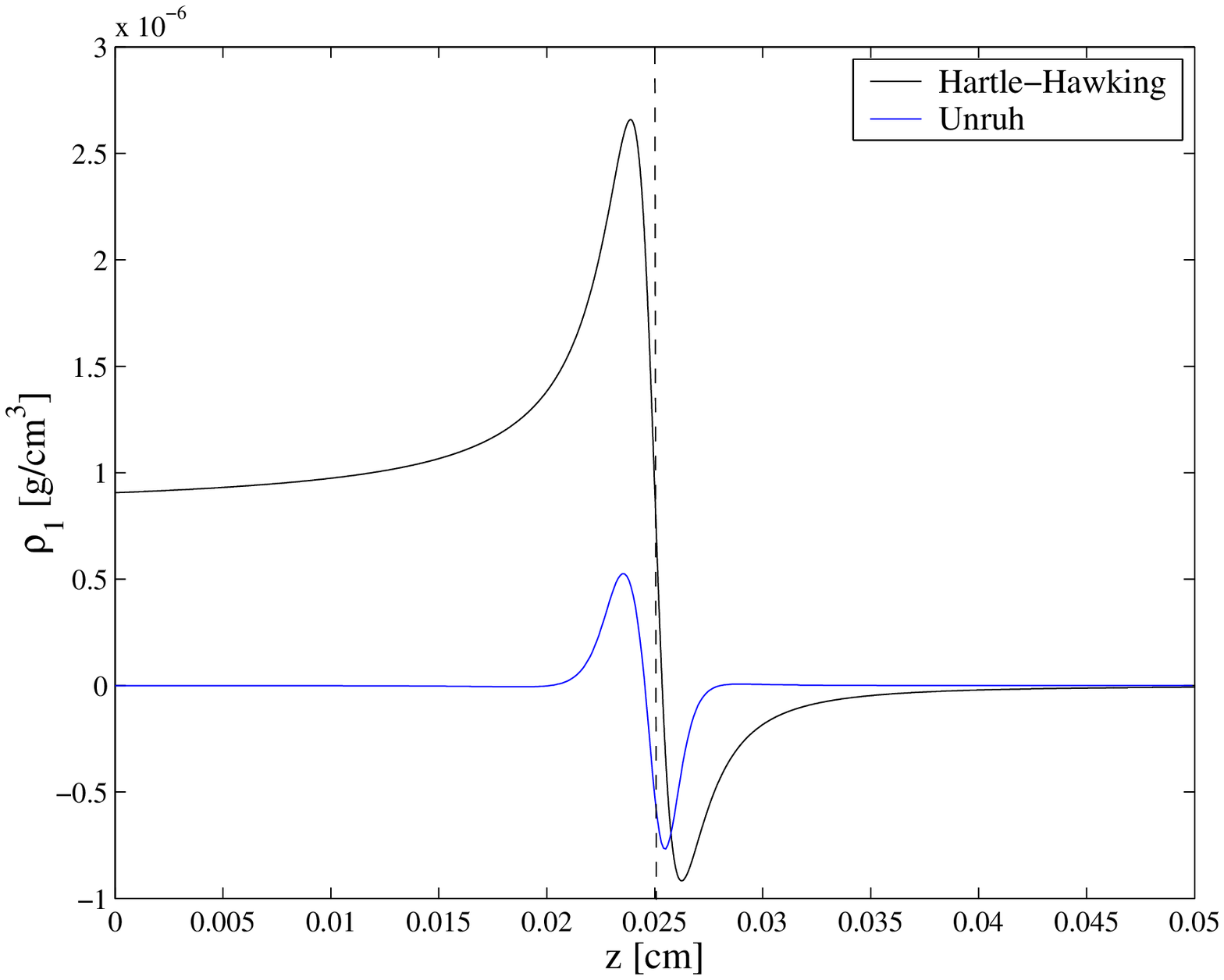} 
\caption{Profile of the quantum correction to the velocity $v_1$ ({\it left panel}) and to $\rho_1$ 
({\it right panel}) due to backreaction in the Hartle-Hawking state (black lines), compared with the profile 
of $v_1$ and $\rho_1$ at $t=t_{\rm end}$ in the Unruh state (blue lines). The vertical dashed lines correspond 
to the position of the sonic horizon at $t=0$.}
\label{label:fig6}
\end{center}
\end{figure*}

Initial conditions are chosen so that at $t=0$ the solution is the classical one; 
i.e., $\rho_1(t=0)=\psi_1(t=0)=0$. Then the backreaction is switched on. As in 
Ref.~\cite{nostri}, since the quantum sources are computed only for the static classical 
background, the validity of the solution is limited by the condition $c\kappa t\ll 1$, 
where we have introduced the constant $\kappa$ as
\beq 
\label{kappa.def}
\kappa = c^{-2}k  = c^{-1}\left. \frac{dv}{dz}\right|_{z_H}
\eeq
with dimension $[\rm length]^{-1}$.
For the sonic black hole considered here, the short time condition determines a 
maximum evolution time ($c\kappa t= 1$) of $t_{\rm max}=0.104\ \mu s$; so it is 
possible to extract only informations about how the backreaction starts. 

Before discussing our numerical results, we briefly describe the numerical algorithms 
implemented, further details can be found in Appendix~\ref{appendixA}.

We are dealing with a system of Partial Differential Equations (PDEs), where the equation for
$\rho_1$ is of convection-diffusion type, due to the parabolic term proportional to $\psi_1''$, 
while the equation for $\psi_1$ is a simple hyperbolic advection equation. As a result, the 
numerical algorithm must be designed accordingly \cite{gustaf}. For the equation for $\psi_1$ a 
simple first-order upwind method is well suited to solve it; for the parabolic equation we have 
implemented standard Forward-Time-Centered-Space (FTCS) explicit method, as well as standard 
Backward-Time-Centered-Space (BTCS) implicit method. Due 
to the short evolution time needed, the limitation of the time--step required by the FTCS and 
the consequent high number of iterations is not a drawback; in any case, we tested one method versus 
the other and we obtained equivalent results. In fact, to have a stable evolution, 
the time step is selected according to the condition $\Delta t=\alpha\Delta z^2/\mbox{max}(\rho)$, 
since $\rho$ is the coefficient of the $\psi_1''$ term in the equation for $\rho_1$. In addition, 
for the nozzle considered, we have checked that a resolution of $\Delta z= 2.5\times 10^{-5}$ cm
(which corresponds to $2000$ points covering the numerical domain) is sufficient to be in 
the convergence regime (see Appendix~\ref{appendixA} for discussion).

In Fig.~\ref{label:fig5} we have snapshots of  the time evolution of the profiles of $v_1$ (left panel) and 
$\rho_1$ (right panel). For this particular computation, we have considered a total evolution 
time $t_{\rm end}=0.2\ t_{\rm max}$. The initial and final snapshots are depicted in red and blue 
respectively. The time delay between one snapshot and the other is $\Delta t\simeq 1.74\times 10^{-3}
\mu $s. The quantum corrected velocity is obtained as $v_1=\partial_z \psi_1$, the derivative 
being computed directly from the numerical data by means of a second order finite-difference approximation.

The numerical solution confirms the near horizon behaviour obtained in Ref.~\cite{nostri}: the fluid 
slows down close to the horizon ($v_1>0$, remember that $v<0$ because the fluid flows from right to left), causing the horizon to move to the 
left, and the total density decreases ($\rho_1<0$). In addition, now (even if for small times) it 
is possible to see the influence of the quantum corrections all over the sonic hole, and not just 
in the neighborhood of the horizon. As a consequence of the shape of the quantum sources $F_2$ and 
$G_2$ (see Fig.~\ref{label:fig4}) the complex structure of Fig.~\ref{label:fig5} emerges. 
One can see  that in the region near the horizon the fluid slows down, but there are also regions where the phonons 
emission induces acceleration. 

\section{Hartle-Hawking  state}
\label{sec:HH}

The thermal equilibrium configuration of the Hartle-Hawking state is much simpler to treat. 
Since the time dependence drops off, the backreaction equations (\ref{continty.back.lin}) and 
(\ref{bernoulli.back.lin}) become a simple system of algebraic equations relating $\rho_1$ and $v_1$:
\begin{align}
\label{continuity.back.HH}
A\left(\rho_1 v+\rho v_1\right)&=\int_{z_H}^z d\xi F_2(\xi)+\mbox{const}\ , \\
A\left( vv_1+\frac{c^2}{\rho×}\rho_1\right)&=G_2 \ .
\end{align}
The integration constant in Eq.~(\ref{continuity.back.HH}) is chosen to be zero in order 
to make the solution non singular on the horizon.

The profile for $v_1$ and $\rho_1$ are depicted in  Fig.~\ref{label:fig6} (black line); for the sake of 
comparison, we show in the same plot the profile of $v_1$ and $\rho_1$  in the Unruh state for $t=t_{\rm end}$
(blue line). In both cases the quantum backreaction correction to the velocity is positive at
$z=z_H$ (vertical dashed line).

In the region very close to the horizon one can make as in Ref.~\cite{nostri} a Taylor expansion 
for the background up to order $O((z-z_H)^5)$. This allows the source terms to be evaluated up to linear term
\bea
F_2&=&\frac{|D|A_H \kappa^3}{96\pi}\bigg[ -(\pi^2+10) \\
   &-&\dfrac{\pi^4+25\pi^2-24}{2}\kappa (z-z_H) +O(\kappa^2 (z-z_H)^2) \bigg]\ ,\nonumber\\
G_2&=&\frac{A_H^2 c^2 \kappa^2}{48\pi}\nonumber\\
   &\times&\left[ (\pi^2+6)\kappa (z-z_H) +O(\kappa^2 (z-z_H)^2) \right]\ .
\eea
The corresponding quantum corrections to the velocity and to the density are
\begin{align}
v_1 &= \frac{A_H c\kappa^2}{192\pi}\bigg[2+\pi^2\nonumber\\
    &-\dfrac{52+35\pi^2+\pi^4}{4}\kappa (z-z_H) + O(\kappa^2(z-z_H)^2)\bigg] \ ,\\
\rho_1 &= \frac{|D|\kappa^2}{192 c\pi}\bigg[2+\pi^2\nonumber\\
       &  +\dfrac{44-19\pi^2-\pi^4}{4}\kappa(z-z_H)+O(\kappa^2(z-z_H)^2)\bigg] \ .  
\end{align}

Setting $v_B=v+\epsilon v_1=-c$
one finds the quantum corrected position of the horizon $z_{H}^q$
\beq\label{Hcorrected}
z_{H}^q=z_H-\frac{\pi^2+2}{192\pi}\,\epsilon A_H \kappa\ ,
\eeq
which is shifted to the left of $z_H$.\\
The quantum corrected equilibrium temperature can also be simply obtained by evaluating
Eq.~(\ref{eq:k}) at $z=z_H^q$ with $v$ replaced by $v_B$. The result  is
\begin{equation}\label{Tcorrected}
T_H^q = \frac{\hbar c\kappa}{2\pi \kappa_B}\bigg[1-\dfrac{\epsilon A_H}{768\pi}\left(52+35\pi^2+\pi^4\right)\kappa^2\bigg] \ ,
\end{equation}
which indicates that, taking into account the backreaction, the equilibrium temperature is lowered.
\section{Conclusions}
\label{sec:conclusions}
Using the continuity and Bernoulli equations, the quantum correction (first order in $\hbar$) 
to a classical stationary flow describing an acoustic black hole has been evaluated in a 
one-dimensional approximation.\\
The quantum corrections to the velocity $v_1$ and to the density $\rho_1$ profiles for the 
equilibrium configuration (Hartle-Hawking state) are depicted in Fig.~\ref{label:fig6} 
(black lines). The phonons backreaction causes the fluid to slow down in the supersonic region, 
with the consequence of a shift of the horizon to the left of the waist of the nozzle 
(see Eq.~(\ref{Hcorrected})). 
In the subsonic region the velocity increases, but the magnitude of the change is smaller than the previous 
one. One finds a similar shape for the density correction $\rho_1$, which increases in the supersonic 
region and slightly decreases in the subsonic one. Finally the equilibrium temperature appears to have 
been lowered by the backreaction from its zero-order value $\hbar c\kappa/2\pi \kappa_B$ 
(see Eq.~(\ref{Tcorrected})).

For the time-dependent case (Unruh state) the analysis has been restricted  to very short times 
after switching on the phonons radiation. This because the quantum source ($F_2$ and $G_2$ in 
Eqs.~(\ref{eq.back.1})-(\ref{eq.back.2})) has been computed only for the classical background, 
which just represents the initial configuration of the acoustic black hole. A more rigorous analysis 
requires the time-dependence of the sources to be included.

Within these limitations, one sees (Fig.~\ref{label:fig5}) a deceleration of the fluid in the supersonic 
region, which causes a drift of the horizon towards the left of the nozzle. Two acceleration regions 
also appear on both sides of the horizon, but the intensity of the effect is lower.
On the other hand, the density correction $\rho_1$ reflects the behaviour it shows in the 
Hartle-Hawking state (on a reduced scale).

\section*{ACKNOWLEDGMENTS}

This work  started when A. N. was visiting  the Department of Astronomy and Astrophysics of 
the University of Valencia. He thanks J.A.~Font, L.~Rezzolla and O.~Zanotti for suggestions 
about the numerical part; in addition, he acknowledges the support of J.~Navarro--Salas and 
the hospitality of the Department of Theoretical Physics of the University of Valencia. 
S.F. acknowledges the Enrico Fermi Center for supporting her research.

\appendix
\section{Numerical schemes}
\label{appendixA}
In this section we report explicitly the time-evolution algorithms. In the main text, we said that 
we used a standard upwind method for the equation for $\psi_1$ and standard explicit Forward-Time-Centered-Space (FTCS) 
or an implicit Backward-Time-Centered-Space (BTCS) schemes for
that for $\rho_1$. In practice the upwind method reads~\cite{leveque}
\begin{align}
\psi_{1,i}^{n+1}&=\psi_{1,i}^n-v_i\frac{\Delta t}{\Delta z}\left(\psi_{1,i+1}^{n}-\psi_{1,i}^n\right) \nonumber\\
&\qquad\quad+\Delta t\left(-\frac{c^2}{\rho_i}\rho_{1,i}^n+G_{2,i}A_i^{-1}\right) \ . 
\end{align}
The FTCS (explicit) and the BTCS (implicit) schemes for the equation for $\rho_1$ respectively read
\begin{widetext}
\begin{align}
\rho_{1,i}^{n+1}&=\rho_{1,i}^n-\frac{v_i\Delta t}{2\Delta z}\left(\rho_{1,i+1}^n-\rho_{1,i-1}^n\right)
+\frac{\rho_i^n v'_i}{v_i}\frac{\Delta t}{2\Delta z}\left(\psi_{1,i+1}^n-\psi_{1,i-1}^n\right)
-\rho_i^n\frac{\Delta t}{\Delta z^2}\left(\psi^n_{1,i+1}-2\psi^n_{1,i}+\psi^n_{1,i-1}\right) \nonumber \\
&+\Delta t\left(-\frac{v_i^2v_i'}{c^2}\rho_{1,i}^n+F_{2,i}A_i^{-1}\right) \ , \\
-\frac{v_i\Delta t}{2\Delta z}\rho_{1,i-1}^{n+1} &+\left(1+\frac{v^2_iv'_i}{c^2}\Delta t\right)\rho_{1,i}^{n+1}
+\frac{v_i\Delta t}{2\Delta z}\rho_{1,i+1}^{n+1} = \rho_{1,i}^{n}+\Delta t\bigg\{F_{2,i}A^{-1}_i+
\dfrac{\rho_i^n v'_i}{v_i}\frac{\psi^{n+1}_{1,i+1}-\psi^{n+1}_{1,i-1}}{2\Delta z}\nonumber \\
   & -\rho_i^n\frac{\psi_{1,i+1}^{n+1}-2\psi_{1,i}^{n+1}+\psi_{1,i-1}^{n+1}}{\Delta z^2}\bigg\} \ .
\end{align}
\end{widetext}
where the cell index $i$ runs from one to $i_{\rm max}$. In the case of the BTCS scheme $\rho_1$ is 
obtained at every time slice (labelled by index $n$) as the solution of a tridiagonal linear system 
of the form $a_{i}u_{i-1}^n+b_iu_i^n+c_iu_{i+1}^n=f_i^n$ that can be accomplished by a standard 
{\it lower-upper} ($LU$) decomposition of the matrix to be inverted~\cite{NR}. A careful treatment 
of the boundaries $z=0$ and $z=z_{\rm c}$ of the numerical domain is crucial for selecting the correct 
solution, especially when the implicit method is employed and so the inversion of the associated coefficient 
matrix is concerned. According to the physical meaning of the Unruh state, we impose outgoing conditions 
at both boundaries, i.e. $u_{i-1}^n=u_i^n$ at $i=1$ and $u_{i+1}^n=u_i^n$ at $i=i_{\rm max}$ where $u^n$ 
can be either $\rho_1$ or $\psi_1$. If $\rho_1$ is solved using the FTCS scheme, since the method is explicit
and no matrix inversion is needed, the problem of setting correct boundary conditions is less important; 
in fact, it is enough to put the boundaries far enough from $z_{H}$ to avoid any influence on the 
evolution. For this kind of equations, stability has also proved to be an issue. Implementing the 
FTCS scheme, to have a stable evolution the time step is selected according to the condition 
$\Delta t=\alpha\Delta z^2/{\rm max}(\rho)$. For the nozzle model discussed in this paper, we have 
used $\alpha=1\times 10^{-4}$ to avoid stability problems and the same choice was kept also for the BTCS
scheme, which results in roughly $3\times 10^{4}$ integration steps. This is not particularly expensive 
from the computational point of view. For example, for the results presented here we used a resolution of
$\Delta z= 2.5\times 10^{-5}$ cm (corresponding to $2000$ grid points) and the total evolution 
took $\sim11$s time evolution on a single-processor machine with a Pentium$^{\rm TM}$  M processor at 1.3GHz. 
The code was compiled using an Intel$^{\rm TM}$ Fortran Compiler. 

We checked convergence of the numerical method (using both BTCS or FTCS schemes) using resolutions 
of 500, 1000, 2000 and 4000 points, considering the case of 8000 points as reference. We computed the 
error $\Delta f$ with respect to the reference resolution as a root mean square for $f=\psi_1$ and $f=\rho_1$. 
From the relation $\Delta f={\cal K}\Delta z^{\sigma}$ we evaluated the convergence rate $\sigma$ and we
obtained $\sigma\approx 1.3$ for both $\psi_1$ and $\rho_1$. We have verified that $2000\div 4000$ grid 
points are sufficient to be in the convergence regime.


\begin{thebibliography}{99}
%
\bibitem{hawking}
	S.W.~Hawking, Nature {\bf 248}, 30 (1974)
%
\bibitem{unruh81}
	W.G.~Unruh, Phys. Rev. Lett. {\bf 46}, 1351 (1981)
%
\bibitem{courant49}
	R.~Courant and K.O.~Friedrichs {\it Supersonic flows and shock waves}, 	Springer-Verlag (1948)
%
\bibitem{libro}
	{\it Artificial black holes}, eds. M.~Novello, M.~Visser and G.E.~Volovik, World Scientific, River Edge, USA (2002)
%
\bibitem{BLV}
  C.~Barcel\`o, S.~Liberati and M.~Visser, {\it Analogue Gravity}, gr-qc/0505065 (2005)
%
\bibitem{nostri}
	R.~Balbinot, S.~Fagnocchi, A.~Fabbri and G.P.~Procopio, Phys. Rev. Lett. {\bf 94}, 161302 (2005);
	R.~Balbinot, S.~Fagnocchi and A.~Fabbri, Phys. Rev.   {\bf D71}, 064019 (2005)
%
\bibitem{unruh76}
	W.G.~Unruh, Phys. Rev.  {\bf D14}, 870 (1976)
%
\bibitem{HH}
	J.B.~Hartle and S.W.~Hawking, Phys. Rev.  {\bf D13}, 2188 (1976)
%
\bibitem{polyakov}
	A.M.~Polyakov, Phys. Lett.  {\bf 103 B}, 207 (1981)
%
\bibitem{sandro-pepe}
	A.~Fabbri and J.~Navarro-Salas, {\it Modeling black hole evaporation},
	Imperial College Press, London  (2005).
\bibitem{york}
For a gravitational black hole the analogous expansion parameter is given by the square of the ratio 
between the Planck length and the horizon size, i.e. $\epsilon=\hbar/M^2$ in units $G=c=1$, see J.W. York, {\it Phys. Rev.} {\bf D31}, 775 (1985)
%
\bibitem{gustaf}
  B.~Gustafsson, H.O.~Kreiss and J.~Oliger, {\it Time dependent problems and difference methods},
  John Wiley and Sons, Inc. (1995)

\bibitem{leveque}
 R.J.~Le~Veque, {\it Numerical Methods for Conservations Laws}, 
 Birkh\"auser Verlag (1992)
%
\bibitem{NR}
 W.H.~Press, S.A.~Teukolsky, W.T.~Vetterling and B.P.~Flannery, {\it Numerical Recipes, 
 The Art of Scientific Computing}, Cambridge University Press (1992)

\end{thebibliography}
\end{document}